# SOFTWARE ENGINEERING STANDARDS FOR EPIDEMIOLOGICAL MODELING

Jack K. Horner
email: jhorner@cybermesa.com

John F. Symons
email: johnfsymons@gmail.com

## 1.0 Introduction

There are many tangled normative and technical questions involved in evaluating the quality of software used in epidemiological simulations. In this paper we answer some of these questions and offer practical guidance to practitioners, funders, scientific journals, and consumers of epidemiological research.  The heart of our paper is a case study in which we provide an analysis of the Imperial College London (ICL) covid-19 simulator.   This simulator is arguably the most influential piece of scientific software in the history of public health policy-making.  Our analysis, combined with reflection on the state of the art in the philosophy of epidemiology and the ethics of engineering serves as the basis for our recommendations for future epidemiological modeling projects. We contend that epidemiological simulators should be engineered and evaluated within the framework of safety-critical standards developed by consensus of the software engineering community for applications such as automotive and aircraft control systems.  To achieve that goal, the development and use of epidemiological simulators must have high levels of transparency, explainability, and reproducibility.  Furthermore, we recommend that such standards be mandated by funding agencies for epidemiological contexts that have direct and significant public policy implications.

We begin by explaining some of the unique features of epidemiological modeling. This discussion draws on recent work in the philosophy of epidemiology. We then highlight relevant research on the epistemology of computational modeling and simulation. From there, we turn to engineering ethics and practice, describing the consensus framework for software engineering that has developed over the past four decades in the software engineering community.  The purpose of this framework is to provide a principled approach to balancing development cost and schedule against the possible harms of using software in high-risk venues.  Within that framework, we evaluate the publicly accessible simulator archive of the Imperial College London (ICL) covid-19 simulator (ICL 2020c).  Our study of ICL 2020c demonstrates that it does not satisfy the standards for safety-critical software in established industry and government practice.

We recognize that ICL 2020c has been subject to intense critical scrutiny because of its role in government decision making during the covid-19 pandemic, and our purpose here is not to pile additional criticism on the work of the ICL team.  In this paper, we focus solely on the publicly available artifacts associated with the simulator (ICL 2020c).   We do not assess the empirical assumptions or epidemiological methodology employed by the ICL team. Instead, we hope that by carefully considering this high-profile epidemiological simulator, we can encourage scientific and philosophical communities to reflect on the norms governing the engineering of scientific software in a wide range of important contexts.

Determining appropriate software standards for epidemiological simulators is not a straightforward matter: it requires an ongoing interdisciplinary effort. In addition to questions of social value and moral responsibility, epidemiological simulators are often informed or constrained by a range of epistemic, mathematical, economic, and technological considerations. Philosophers of science are beginning to understand the trade-offs among these factors and are becoming increasingly sensitive to the implications of software-intensive scientific inquiry (Symons and Horner 2014). Philosophers have also recognized that we cannot understand appropriate norms for scientific practice solely by reasoning about them a priori - we must take empirical evidence and technical constraints into account.[1] The standards governing the development of important epidemiological simulators cannot be left solely to epidemiological practitioners. The development of these standards requires careful normative reasoning that is often beyond the expertise of such practitioners. In addition to falling outside of their area of scientific specialization, the normative questions governing scientific inquiry involve balancing the interests of the practitioners themselves with those of others. Standards for the development of epidemiological simulators must accordingly accommodate the interests of both the consumers of epidemiological research and the interests of the broader public that is affected by the public policy decisions influenced by this research. In the fraught context of epidemiology this task involves balancing competing social values.

## 2.0 Policy making, social values, and epidemiological computer simulations

Computing technologies have played an important role in medical and biological practice in economically developed societies since at least the 1960s (See e.g. Keller 2002; November 2014). Evaluating the role and usefulness of data-driven computational models and simulations is complicated in biological contexts for reasons others have explored in detail (See eg. Leonelli 2011; 2012; 2016, and Stevens 2017 for example). Epidemiology is an even more challenging context for evaluating the role of computational models and simulations than, for example, in molecular biology for reasons we will discuss below. More generally, as many philosophers have noted, computational models have distinctive technical and epistemological features that make them uniquely difficult to assess.[2] While philosophers have addressed the role of data science and computational modeling in biological contexts for two decades there has been relatively little scholarly attention given to the norms governing the engineering practices underlying these models. Likewise, while philosophers of epidemiology have correctly emphasized the normative and political aspects of research in epidemiology, they have largely neglected the norms guiding engineering practices in the development of epidemiological models and simulations. Since models and simulations are fundamental to epidemiological predictions and recommendations, they should also be subject to critical scrutiny.

---

[1] Developers of any large piece of software must face practical and theoretical constraints on error correction. See, for example, Horner and Symons 2019, Symons and Horner 2019.
[2] See for example López-Rubio and Ratti (2019) for a discussion of the trade-offs between mechanistic explanation and prediction in applications of machine learning to molecular biology.

In the context of epidemiology, computational modeling and simulation techniques have become indispensable research tools (See Smolinski et. al 2003). As Boschetti and colleagues have noted, computational modeling is sometimes our only way of advancing scientific inquiry in contexts where ethical considerations or practical constraints prevent the use of traditional experimental techniques (Boschetti et al 2012). In epidemiology, highly complicated simulations and the manipulation of large data sets, along with the ethical and practical obstacles to experimentation have meant that computational methods have become centrally important research tools.

During the COVID-19 pandemic in particular it was widely reported that the results of computer simulations provided by the ICL team weighed decisively in public policy deliberations in both the UK and US governments (Landler and Castle 2020). Government officials are reported to have relied on epidemiological modeling and simulation to predict mortality due to the virus and to anticipate its effects on the healthcare system. These simulations are also used to assess the relative merit of alternative interventions and public health responses to the pandemic (Freedman 2020).[3] In an emergency decision making context, it is reasonable to turn to acknowledged experts on the relevant topics and throughout the pandemic, political decision makers in the UK government have been eager to present their decisions as grounded in the best available scientific evidence (UK Government 2020). The extent to which decision makers have or have not 'followed the science' has become a fraught and highly politicized matter in many democratic societies (Stevens 2020; Sharma 2020).

The role and status of computer simulations frequently figures, albeit unsystematically, in debates about what it means for governments and institutions to 'follow the science'. It is clear from reporting and from the actions of the UK government that the simulation results provided by the ICL team were decisively important in policy deliberations in March and April of 2020 (Landler and Castle 2020).[4] It is also clear that the ICL group occupied a high position of scientific authority and trust from the perspective of political decision makers.

In these discussions it is often mistakenly assumed that policy is fully determined by our best epidemiology. As we explain below, this assumption involves a misunderstanding of both the nature of epidemiology and its proper role in decision-making in democratic societies. Difficult trade-offs between different kinds of societal values and moral obligations will not, generally, be resolved by scientific expertise. Epidemiologists cannot tell us, for example, in the case of the COVID-19 epidemic whether public health interventions ought to value the well-being or education of children more highly than reducing the health risks to the elderly. These are moral and political decisions that are not illuminated directly by increased scientific understanding or better models and simulations.

'Follow the science' presumably means 'follow the *best* science'. However, determining which epidemiological recommendation is best is not a straightforward matter. Given the complexity of the factors relevant to decision making during a pandemic, the public in democratic societies and their political representatives have placed great trust in the community of epidemiologists. This is understandable, but often, public declarations of trust have implicitly projected an idealized and unrealistic level of neutrality and objectivity onto epidemiological research. This runs counter to our best critical understanding as drawn from philosophy of epidemiology. As we will explain below, disagreement among epidemiologists often stems from differences with respect to values (Stevens 2020).[5] The way we have assigned trust to the epidemiological community is not unreasonable, but it involves oversimplification that can lead us to misunderstand our responsibilities as consumers of their research. We are operating with something like the following commonsense understanding of the relationship between scientific expertise and policy making:

> *Commonsense view of scientific evidence as a guide to policy making*
>
> Decisions that involve risk of serious harm require us to deliberate as carefully as is feasible. Policy makers often have to rely on expert advice since our best available evidence and guidance for decision-making in many matters comes from scientific experts. In such contexts, it is usually rational to follow the advice of the relevant scientific community in order to increase the likelihood that our decisions promote our values and interests. Commonsense recognizes that natural science cannot tell us what we ought to value or what our policy goals ought to be. Nevertheless, under ideal circumstances science can provide an understanding of the facts in a way that helps us to act consistently with the moral or political principles we share.

---

[3] Perhaps the most important policy role of these simulations has been their perceived predictive power. For a discussion of the predictive role of computational models see Boschetti and Symons (2011) and Symons and Boschetti (2013). See Ioannidis et al. (2020) for an assessment of the predictive power of prominent covid-19 modeling efforts to date.

[4] In mid-March, the ICL model was predicting that absent any public health interventions, the UK would suffer half a million deaths from COVID-19 (2 million deaths in the U.S.).

Much in this prescription is in our view correct. However, it draws a sharp distinction between social values and norms and scientific inquiry in a way that is clearly inappropriate in the case of epidemiology for reasons we will explain below. The assumption that epidemiology is value neutral makes our insistence on the importance of high standards for scientific software seem like an unwarranted intrusion on scientific practice. However, there are degrees of neutrality in the sciences when it comes to values. For example, when one turns to an epidemiologist for advice, one cannot be as confident of the value neutral nature of their scientific judgment as one would be in discussions with a chemist. Philosophy of epidemiology has highlighted the complex moral and political landscape of the study of epidemics. In this context the standards governing how software for simulations ought to be applied are similarly complex.

Disputes within epidemiology involve normative considerations in ways that disputes between chemists or physicists, for example, almost never do. Consider debates concerning the social determinants of health, where disputants may offer causal stories about the origins of some public health concern that assume, or are motivated at least in part by their preferred socio-political values.[6]  In epidemiology, social, political, and other considerations are difficult to disentangle from the manner in which scientific questions are framed.  The way epidemiologists think about causation, agency, possible interventions, relevant populations, risk, disease, and responsibility, are all informed by the values governing their practice.[7]

---

[5] See Stevens (2020) for a discussion of the confusion around 'follow the science' rhetoric in UK policy making. He writes: "A provisional and contested set of statements about how the world is cannot be used directly as a rule for what governments should do. Ministers have to decide for themselves. They must take responsibility for these decisions and their own inevitable mistakes, rather than relying on science as if it were an apolitical and indisputable tablet of stone." https://doi.org/10.1038/s41562-020-0894-x

Public reflection on norms is relevant for the practice and not just the application of epidemiology. In order to explain why, consider a disease like type-2 diabetes. There are interventions that would be effective in stopping the spread of this disease that we would regard as unconscionable violations of individual autonomy, or that most of us would presently regard as contrary to the ultimate goals of public health.[8]  For example, we might reject heavy taxation on calorie dense foods, mandatory exercise programs, etc. as possible responses to the disease because of the importance of other kinds of social goods. Generally speaking, the set of acceptable interventions available to us for public health problems will be shaped by a range of social values.

In addition to disagreeing with respect to what would count as an acceptable intervention in public health, social groups may also disagree over what kinds of health issues should be classified as diseases or as epidemics.  There is considerable disagreement over, for example, the claim that common mental health problems like anxiety and depression should be regarded as epidemics.[9] Claims that obesity or attention deficit hyperactivity disorder are at epidemic levels in the United States, for example, are difficult to state categorically without reference to a large set of controversial normative assumptions. Ultimately, social values are negotiable. People with differing values can attempt to persuade one another with respect to the relative importance of conflicting values.  Given the role of social values in determining the space of acceptable public health interventions, the characterization of health, and the taxonomy of disease, epidemiological inquiry will always be situated within a particular social context and cannot be entirely neutral with respect to normative questions.

Attempts to characterize the subject matter of epidemiology will also generally require reference to concepts that have normative features. Mathilde Frérot and colleagues surveyed the literature from 1978 to 2017 in order to determine the ways that epidemiologists understand their enterprise and how that understanding has changed through time.  They examine 102 definitions of 'epidemiology' and found that five terms were present in more than 50% of definitions: "population", "study", "disease", "health" and "distribution" (Frérot et al. 2018).  Philosophers of epidemiology have noted that definitions of epidemiology will vary depending on the social and political contexts involved.  In their introduction to the recent *Synthese* volume on philosophy of epidemiology, Jonathan Kaplan and Sean Valles emphasize this contested nature of epidemiology (Kaplan and Valles 2019).[10] They contend that "since the welfare of populations and communities are always at stake in epidemiology, the issues at hand are directly or indirectly political issues" (Kaplan and Velles 2019).

---

[6] See Broadbent 2012 for a discussion of causal reasoning in epidemiology
[7] As mentioned above, for example, during the early stages of the COVID-19 crisis, harm to the education of the young and risks to the life of very elderly people were weighed against one another without a great deal of explicit public deliberation.  The assumptions about social priorities that motivated school closures and other interventions that harmed children and young people may well be defensible.  The kinds of interventions that were attempted in the early stages of the pandemic are all defensible given *some* set of social values.  In most cases, epidemiologists did not engage in explicit and public deliberation concerning their presuppositions about social values when they offered their initial recommendations with respect to interventions.
[8] See Tabish et.al 2007 for a defense of categorizing diabetes as an epidemic.
[9] See Baxter et al 2014 for an argument against considering common mental health problems like anxiety and depression as epidemics.

Our task in this paper is to encourage attention to the norms governing software engineering in epidemiology. The significance of these models for policy decisions that affect many of us in significant ways is clear. Given that epidemiology is not normatively neutral to the degree that scientific disciplines like chemistry or physics might be, there are good reasons to believe that non-practitioners have a right and an interest to concern themselves with the standards governing software engineering in this discipline. Funders, journals, policy makers, and the broader public are entitled to require that standards are sufficiently high to ensure that simulations are trustworthy. In the next section we discuss some of the most important aspects of trustworthiness for computer simulations. As we shall argue, part of determining the standards for what count as good software engineering practice will be determined by the level of risk involved in the deployment of the simulation.

## 3.0  The epistemology of epidemiological computer simulations

As discussed above, there has been significant public interest in the trustworthiness of epidemiological models. Most criticisms have raised doubts concerning the assumptions and the quality of the data that go into the models rather than with respect to the quality of the software underlying simulations. Our focus in the following is on properties of the software as software, rather than the scientific status of the assumptions, the mathematical models, or the quality of the data driving these simulations. Occasionally, critics have pointed to weaknesses in the publicly available code for simulators and we will address some of these criticisms below (Lewis 2020).[11] As we argued in the previous section, the quality of the engineering, in addition to the quality of the science is a relevant topic for philosophical reflection.

In recent years philosophers have addressed some of the central epistemic problems associated with computer simulations in science.[12] One standard approach to understanding *why* scientific communities come to trust simulations relies on an analogy with the ways that epistemic entitlements work in other less controversial forms of inquiry (See for example Barberousse and Vorms, 2014). In ordinary life, for example, we are generally entitled to trust the testimony of other people, the reliability of our senses, and the capacity of our basic cognitive faculties, such as our memory to transmit information without altering it in epistemically significant ways. This use of the idea of epistemic entitlement, largely drawn from Tyler Burge's (1993; 1998) arguments, has been highly influential among philosophers in debates over the epistemology of computer simulation (Barberrouse and Vorms (2014), Beisbart (2017)). Symons and Alvarado disagree, arguing instead that the analogical account of epistemic warrants is not appropriate in the context of computer simulations. They have insisted instead on epistemic standards of the kind we apply to traditional scientific instruments (Symons and Alvarado 2019; Alvarado 2020). On this view, computer simulations are not experts and should not be treated as such. Instead, they are built by teams of experts or by experts working alone who may not be expert software engineers. Thus, given the interdisciplinary integration necessary in a team, the use of the analogy with trusting experts is inappropriate. The analogy is even less fitting in the specific case of epidemiological simulation than it is in science more generally, given that epidemiology relies on interdisciplinary teams with distinct sets of disciplinary standards. Furthermore, the resulting simulations are heavily mediated by what Eric Winsberg called motley practices (Winsberg 2010).

---

[10]They contrast what they see as the divergence between the views of the World Health Organization and the United States Centers for Disease Control. While they do not provide evidence for divergence between these two organizations, they do note two conflicting characterizations of epidemiology, both of which are drawn from the Dicker et. al's (2006) document for the CDC: "Epidemiology is the study of the distribution and determinants of health-related states or events in specified populations, and the application of this study to the control of health problems." (2006, I-1) and later in the same document: "in epidemiology, the 'patient' is the community" (2006, I–4). See Frérot et al. 2018 for a careful empirical assessment of the variety of ways that epidemiology has been characterized in recent decades.

[11] There has been considerable popular attention to the issue of the quality of code in epidemiological simulations during the COVID-19 pandemic. The quality of these analyses is mixed, for a flavor of some of the commentary see for example Lewis 2020.

[12] See Juan Duran (2018) and Winsberg (2019) for an overview of the epistemic issues related to computer simulation.

The fact that we are not able to trust computer simulations by analogy with the manner in which we trust individual scientific experts leaves us with the problem of how policy makers and the public should decide which simulations and which models to rely upon. There are many dimensions to this challenge and it is beyond the scope of this paper to address this broader problem (see, for example, Symons and Alvarado 2019). Trusting simulations involves many complicated criteria. However, for the remainder of this paper we will argue that at least one obvious and necessary condition for justifiable use of simulations for public policy is that they be funded, managed, specified, designed, implemented, and maintained in accordance with the best available software engineering practices. We contend in this paper that these practices are as important to software-intensive policy-making as good experimental methods are in non-software-intensive scientific regimes. Our recommendations will increase the cost of these simulation efforts and will require increased collaboration between scientists and software engineers. However, we contend that the risks involved in decisions based on epidemiological modeling efforts warrant the additional resources and effort that we recommend here.

## 4.0 Standards for software engineering

As with all aspects of epidemiology, engineering standards governing the development of simulations are a matter where normative considerations overlap with technical and mathematical constraints. Because of this, critical scrutiny of these simulations is not the exclusive purview of any subset of scientific experts as we argued above. Practical guidelines for developers of scientific software are described and defended below. We did not invent these standards. Instead, our recommendations draw upon the recent history of engineering. For the past five decades, the software engineering community has sought to codify practices and procedures that have been empirically determined to help minimize development cost, schedule, and risk (Boehm 1973; Myers 1976; Boehm, Abts, Brown, Chulani, Clark, Horowitz et al. 2000). This effort has produced an evolving series of software engineering standards, one of the most recent of which is ISO 2017.

We should begin by acknowledging that scientific software is generally not developed according to standards anywhere near as demanding as ISO 2017. Thus, our recommendations will be controversial and may be regarded as excessively restrictive for those who view epidemiology solely in terms of its scientific character.

The most controversial feature of our proposal is the application of an approach drawn from engineering ethics to a discipline that primarily regards itself as a science.[13] While the responsible practice of engineering is generally sensitive to the harms involved in various projects (Roddis1993; Lynch & Kline 2000), our critics may respond that a science like epidemiology is different. The ethics of scientific inquiry, one might argue, are very different from the ethics of engineering. Most philosophers are likely to agree. The kinds of simulations that epidemiologists have produced have been regarded by philosophers, for example, as either formal or abstract objects or as special forms of experiments capable of yielding empirical information about the systems they simulate.[14] Following Alvarado (2020) we believe that in addition to serving as formal models or experiments, simulations should also be understood as scientific instruments.

In the pages that follow, we will defend our contention that epidemiological models should be evaluated in the same way we evaluate engineered instruments. As Roddis (1993) notes, in engineering ethics, the standards governing instruments and practices are determined, at least in part, by the harms that can result from failures. We contend that high standards are required for software engineering in epidemiological simulators given the high costs of failure involved in the deployment of these instruments in public policy decision making.

We argue that where great harms can result, scientists, funding agencies, and governments ought to adopt standards of software engineering that are at least as high as the standards that societies routinely demand in, for example, critical infrastructure, aviation, or military contexts. Because of the nature and extent of the harm to societies that errors in these simulations can cause, epidemiological modelers are subject to a special obligation to adhere to high standards in the development of their software.

The principal objection to insisting on such standards is the risk that convergence to a single set of standards might inhibit or slow the development of scientific inquiry. We believe that this risk is overstated and that open and transparent scientific software built to high standards is likely to help rather than hinder the scientific enterprise. Additionally, there is a difference between the kind of anarchic creativity that might permit scientific insight and the construction of scientific instruments like computer simulations.[15]

---

[13] Epidemiologists sometimes present their work as a basic science for clinical practice in medicine. See eg. Sackett et. al (1985) and Bonita et.al (2006)

[14] Weisberg (2012) and Pincock (2011), regard computer models as formal extension of mathematical representation. Morrison, (2009; 2015) regards computer simulations as being a form of scientific experimentation (Ruphy, 2015). Morrison and others have argued that computer simulations involve extra-mathematical considerations (Winsberg, 2018). These include measurement practices (Morrison, 2009), representations and imaging (Barberousse et al., 2009), and hypothesis testing and generation (Hartmann, 2005). Alvarado offers an alternative view of computer simulations as instruments (2020).

[15] For a defense of treating computer simulations as instruments see Alvarado 2020

**5.0  Software engineering standards in pandemic policy-making**

Since the late 1960s, the software engineering community has sought to codify consensus software development practices and procedures that have been (empirically) determined to help minimize development cost, risk (both developmental and operational),  and to help ensure that the products of such projects reflect user needs and values (Boehm 1973; Myers 1976; Boehm, Abts, Brown, Chulani, Clark, Horowitz et al. 2000). These codification efforts have produced a series of software engineering standards.[16]  As mentioned above, one of the most recent and widely used software engineering standards is ISO 2017.  Although there is some variation among these standards, they characterize software projects in terms of lifecycle phases, each with formal review and documentation requirements, both which directly contribute to the transparency, explainability, and reproducibility of the software developed under those standards. These phases are:

1. specification
2. logical design
3. physical design
4. implementation
5. test
6. maintenance

The economic and risk-management rationale for a phase-structured approach to software development and management is based on two major premises (Boehm 1981, 38):

I.   In order to create a "successful" software product, we must, in effect, execute all of the phases at some stage anyway.
II.  Any different ordering of the phases will produce a less successful software product.

Rationale (I) follows directly from questions that inevitably arise in the development of any software system: "What is the software supposed to do?" (Specification phase), "How do we ensure that everyone who helps to develop part the software understands how his/her part of the software correctly integrates with the rest of the software?", especially if not all personnel know all aspects of the system  (Logical, and Physical, design phases),[17] and "How do we determine that the software is doing what is supposed to do" (Test phase).

   Rationale (II) derives directly from empirical studies of the costs of fixing an error in a software system as a function of the phase in which the error is detected and corrected.  These studies show that in a large (> ~50,000 source lines of code (SLOC; Boehm, Abts, Brown,

---

[16] Such a standard is not a contract; in the absence of a contract, compliance with a standard is therefore voluntary. A contract, however, can make compliance with a standard mandatory.

[17] On average, five years after initial deployment of a software system, only 20% the original developers of the software remain on the project ( Boehm, Abts, Brown, Chulani, Clark, Horowitz et al. 2000, 48).  10 years after initial deployment, on average, none of the original developers remain on the project.  On small projects, furthermore, the loss of even a single key team member can force the project to restart or be abandoned.  Detailed documentation is the only way to mitigate these risks.

Chulani, Clark, Horowitz et al. 2000, 395)) or highly technical software project, a typical error is 100 times more expensive to correct in the maintenance phase than in the specification phase; in small projects ($< \sim$10,000 SLOC), a typical error is 20 times more expensive to correct in the maintenance phase than in the specification phase (Boehm 1976; Boehm 1981, 40).

Each of phases 1-6 imposes requirements on, or equivalently, allocates requirements to the processes and products of one or more successor phases. Taken end-to-end, the resulting requirements-allocation induces a hypergraph (Berge 1973) spanning the elements (documentation, processes, and code) in the system.

Documentation is crucial to ensuring the transparency, explainability, and reproducibility of software. It is sometimes incorrectly assumed that the code in a software system determines what that software is intended to do. One reason that the code as such does not fully determine the intended application semantics of the code is this: Any program, regardless of what the code *seems* to be about, could be used solely to show that the machine on which it runs will in some sense cycle the program, without regard to anything else that program is supposed to do. Here are two examples of software, each of which appears to have a definite use based on a reading of the code proper, has a quite different use. In the early 1980s, a large US military data-communications system contained a program, that appeared to be data-communications code (and in fact had once been used as such). By 1985, however, this program was used solely to stress-test the ageing, ailing disk drives in the system. Similarly, in the early 1990s a large European research institute owned what looked like (and in fact once was) a nuclear reactor control code. But by 1995, that code was used solely to test the performance of executables produced by Fortran compilers. The reason for mentioning these cases is that software engineering concerns more than just code. Only the combination of the specification, the logical design documentation, the physical design documentation, various test suites, and the code proper, can capture what the code is supposed to do.

Obviously, there is no guarantee that using a software development process of the kind described in this section will produce an error-free software system.[18] It is all but certain, however, that if such a framework is not used, the system, with very high probability, will contain errors that could have been avoided if such a framework had been adopted (Boehm 1973; Boehm 1976; Myers 1976; Boehm 1981, 40).

The standards allow tailoring or editing, as a function of cost, schedule, risk to property and life, and other harms of comparable consequence. Software whose failure would result in inconsequential loss of property, life, or revenue, for example software developed solely for personal use, can be developed with little formality according to these standards. In contrast, consensus standards in software engineering require that software whose failure could result in large loss of property or life (e.g., aircraft or automobile control) ought to be developed with extensive formality.[19]

While informal software development is often tolerated in academic contexts, standards must be higher in the case of epidemiological modeling that is used in public-health policy-making. Why? The epidemiological simulators used in policy-making are typically used in a way that errors in those simulators could lead to substantial loss of property or life, or to other harms of comparable consequence.

---

[18] See Horner and Symons 2019b for a discussion of whether it is even possible, in all cases of interest, to determine that we have produced error-free software.

[19] For further information, see Boehm, Abts, Brown, Chulani, Clark, Horowitz et al. 2000, Hatton 1995, ISO 2017, Koopman 2014, MISRA 2008, NASA 2004, Rierson 2013, RTCA 2012, and FDA 2002.

The development of general software engineering standards has combined a recognition of both general principles of engineering ethics and attention to the empirical features of software engineering practice. In the next section we apply these standards to ICL 2020c.

## 5.1 A case study

As mentioned above, the Imperial College London (ICL) covid-19 simulator is arguably one of the most influential pieces of scientific software in the history of public health policy-making. Given its role in informing the responses to the COVID-19 pandemic by policy makers in the United Kingdom and the United States it should be developed and managed with the formality required by the standards like ISO 2017.

During the period from late-March through late-May 2020, we assessed how well the publicly accessible artifacts (ICL 2020c) of the ICL covid-19 simulator project conform to the consensus software engineering standards framework outlined in the previous section. As mentioned above, our assessment was based on informed software engineering judgment, reading those artifacts, building and executing some of the code, and applying various analysis tools (identified below) to the artifacts in that archive.

Our assessment was constrained by some important limitations. Most importantly, to our knowledge, there is no publicly accessible documentation that officially identifies the baseline for the ICL covid-19 simulator project, though a cursory inspection of publicly available materials might suggest otherwise. For example, as of 12 June 2020, an ICL covid-19 project website (ICL 2020d) appears to identify the mapping between certain code archives and various team papers and reports. Our analysis revealed, however, that the code archives identified on this website contained modification date/time stamps that are *later* than the issuance dates of these papers and reports. We further discovered that some of the graphics that appeared in the papers and reports referenced on the website were not directly produced by any of the code in the associated code archives. (It is possible, of course, that some of these graphics were produced by applying software that is not identified in the reports/papers or on the website to the outputs of code that does appear in the archives.) In addition, according to Eglen (2020), ICL 2020c is not identical to the version of the code that produced the tables in ICL 2020b ("Report 9"). However, Eglen reports that a CODECHECK assessment of ICL 2020c produces results that agree with the content of some tables in ICL 2020b for the test cases run in Eglen 2020. It is therefore not possible to infer from this website, or from the papers/reports linked at this website, the identity of the specific code used produce the results reported in associated papers and reports.

We note that there is, at present, no legal or institutional requirement for the ICL simulator project to make any software-development artifact of that project accessible to the general public. It is not surprising, therefore, that, even if they exist, many of the artifacts identified in the consensus software engineering standards are not publicly available in the ICL simulator project. In our judgement, however, it is highly likely that ICL 2020c is closer to the actual ICL covid-19 simulator project baseline than any other publicly available artifact; accordingly, we chose ICL 2020c, along with the published articles and reports identified in ICL 2020d, as the baseline for the analysis reported here.

Our first recommendation is that in future epidemiological simulator projects, funding agencies and scientific journals ought to require the provision of a project baseline for the purposes of reproducibility and verification.

Assuming ICL 2020c as the baseline for our assessment, Sections 5.1.1 – 5.1.6 describe, at a high level, the major features of each phase of the software engineering process described in ISO 2017 and assess how well, within the limitations described above, ICL 2020c conforms to that standard. Detail on each of these phases can be found in ISO 2017. For the purposes of this paper, it is sufficient to sketch them only in outline.

### 5.1.1 Specification Phase

The principal function of the specification phase of a software project is to generate an agreement (called the *specification*) among stakeholders that states what objectives a software system must achieve. Among other things, the specification is intended to reflect the results of the negotiation of stakeholder values. In the case of epidemiology simulator development projects, such tradeoffs can concern negotiations of the tradeoffs between the rights of the younger and the elderly, or tradeoffs in optimizing on the social-distancing directives/guidelines collides directly with other activities that all but require person-to-person physical contact. (In several stakeholder communities, these tradeoffs (as of mid-2020) have yet to be resolved.) In some policy-making venues, furthermore, the general public is a stakeholder and thus can legitimately claim a right to have, in a timely way, access to all policy-related artifacts such as simulator rationale, design, and implementation (a view institutionalized, for example, in UK Government Office for Science 2010):

> 73. SACs [Scientific Advisory Committees] and their secretariats should aim to prepare papers in accessible language. Where issues require technical discussion, consideration should be given to separate, and additional, production of a 'lay summary' to ensure that all matters are accessible to all interested parties regardless of specialist knowledge. (UK Government Office for Science 2011, 18)

Justifiable decision-making typically requires transparency and explainability – even insuring, in some cases, some level of lay understanding.[20] Policy makers cannot be expected to be able to evaluate models and simulations at the level of technical detail, but modelers should be transparent with respect to, for example, the degrees of uncertainty involved in their predictions. In complex decision-making problems facing policy makers, modelers must therefore represent the extent to which their predictions should be believed. Trusting experts is unavoidable and fully appropriate in certain domains, especially those with high technical content. However, as we discussed above, expertise in technical, scientific, or engineering domains (such as epidemiology), does not imply expertise with respect to societal goals and values.

---

[20] European Union law establishes a right to explanation in relation to the use of technology in important decisions affecting individual citizens. See for example https://eur-lex.europa.eu/legal-content/EN/TXT/?qid=1465452422595&uri=CELEX:32016R0679 Rectital 71 (accessed June 8 2020). French national law establishes the right to explanation in the 2016 *Loi pour une République numérique*. See also Morely, Cowls, Taddeo, and Floridi 2020. Such a right is not categorical, however. In the case of code and documentation that contains information whose disclosure could compromise national security, access to these artifacts must be restricted.

Relevant value considerations and assumptions shaping the development of the simulation should be explicitly stated in the Specification to the extent possible. The degree to which precautionary or other values enter into the choice of parameters, data sources, etc. should also be captured in the specification, because they can affect our understanding of the meaning of the predictions derived from the simulation.

Unfortunately, there is no publicly accessible specification for ICL 2020c. Ideally, future iterations of this and related simulators ought to be generated according to publicly negotiated specifications. At the very least, the specifications stipulated by the modelers themselves should be made available to the public.

Modelers and their funding organizations might protest that epidemiological simulation is a time-sensitive project whose urgency precludes such public deliberation. We contend, however, that the trust invested in epidemiologists by the public and their political representatives in these contexts means that they must be able to provide a well-articulated and understandable specification. We recognize that maintaining and managing any publicly accessible archive requires non-trivial cost and schedule commitments. A clear specification will explain the purpose and assumptions of the simulator in ways that will help ensure its trustworthiness and will permit all stakeholders to properly evaluate its relevance to their decisions.

### 5.1.2 Logical Design Phase

The objective of the logical design phase is to generate an abstract description, called a Logical Design Document, of a system that satisfies the requirements of the specification. Understanding what "satisfaction" means in the software development process is not simple and it involves considerations beyond the scope of this paper. For an explanation of the notion of satisfaction in the context of software development projects, see Symons and Horner 2019.

The abstract description that satisfies the specification assumes no particular implementation in hardware, software, or human procedures. Various languages can be used to express the logical design. In current practice, the Unified Modeling Language (see, for example, Rumbaugh, Jacobson, and Booch 1999) is often used for this purpose. No software is generated during this phase. There is no publicly accessible Logical Design document in ICL 2020c. This is not unusual for scientific software, but it does violate the consensus standards for software deployed in high risk contexts.

### 5.1.3 Physical Design Phase

The objective of the physical design phase is to generate a concrete description, typically called the Physical Design Document, or Detailed Physical Design Document, of how specific machines, software, and human processes, and their interactions, will satisfy the requirements allocated to them from prior phases. The software-specific component of the Physical Design Document is often called the Software Design Document, or SDD. (For a detailed description of an SDD, see US Department of Defense 1988.) Assuming ~50 software statements per page of source code, this document typically contains ~10 pages per page of source code. No software is generated is generated during this phase. There is no publicly accessible SDD for ICL 2020c. However, a few items that would be contained in an SDD are included in the inline comments of the source code in ICL 2020c.

### 5.1.4 Implementation Phase

This phase implements on actual machines, and in software and human procedures, an operational product that satisfies the requirements allocated to it from prior phases. The software developed during implementation phase is typically required to satisfy certain programming-language-specific standards (sometimes called "coding guidelines"). These standards prescribe programming-language-specific practices that are, and proscribe practices that are not, acceptable. (Such requirements are often stated in, and inherited by allocation from, the specification.) The primary role of these programming-language-specific standards is to minimize programming-language-specific coding errors.[21]

By "manually" analyzing ICL 2020c along with the reports and papers associated with that archive we determined that the source code in ICL 2020c was intended primarily to study the effect of "interventions" (e.g., school closings, social distancing) and population-distribution details on the course of a pandemic. Based on our analysis of inline comments in the source code, and on the style of the code itself, the code in ICL 2020c appears to have descended from a multi-thousand-statement simulator written in the C language by one developer in the early 2000s. In its current form, the code is almost entirely implemented in the C language subset of C++. For example, ICL 2020c makes no use of C++ classes or type polymorphism.[22]

By applying the static source code analyzer *Understand* (Scientific Tools 2020) to the source code in ICL 2020c we found that the code consists of ~1000 declarative/definitional, and ~10,000 executable, statements, distributed across approximately 30 files. Half of these statements are in a single file that contains the source for the simulator's main routine.

The complexity of software serves as a rough measure of the intelligibility and the maintainability of the code (Symons and Horner 2014). All else being equal, software engineering attempts to minimize the complexity of a software system while satisfying all other requirements on that system. There are many way to measure software complexity. One of the more widely used measures of software complexity is *McCabe complexity*. Informally put, McCabe complexity is the number of distinct execution paths through the code (McCabe 1974). Statistically, the frequency of errors in software is an increasing function of McCabe complexity (Basili and Perricone 1984). ~50% of ICL2020c has extremely high McCable complexity. Most of this complexity comes from deeply nested "if, then" statements, the understanding of which requires the reader or software developer of the code to maintain awareness of long chains of conditionality.

---

[21] For examples of such standards, see Hatton 1995; Evans 2003; Perforce 2013; Google 2020.
[22] We made this assessment by reading the ICL 2020c source code, and by analyzing the ICL 2020c source code with the documentation tool *doxygen* (van Heesch 2020) and the static source code analyzer *Understand* (Scientific Tools 2020).

A simulator typically requires that the user enter input values for the parameters that are relevant to the model underlying the simulation. "Manual" analysis of the ICL 2020c source code and its input files reveals that in order to generate a simulation, one must enter 40-50 distinct values. To configure ICL 2020c requires that users have reliable data supporting 40-50 input-variables and parameter-assignments. For the most part these are data derived from public health sources but in some cases it is less clear how these assignments are determined. The high number of parameters in this simulator and its resulting complexity cannot be avoided at some level if the model is to assess the effects of even the intervention regimes that have already been deployed by various countries. As a result, ICL 2020c is unavoidably more difficult to comprehend, correctly use (arguably, only the authors of the code can reliably use it), calibrate, and maintain than lower-fidelity epidemiological models such as SEIR (for a description of SEIR, see Vynnycky and White 2010; Nowak and May 2000). Transparency with respect to these parameters is important in order to ensure the trustworthiness of the simulator.

With the exception of the high-complexity portion of the code mentioned above, the ICL simulator is, as of 15 September 202, being modified in a way that is generally in accordance with at least some of the software engineering standards we have described here. The scope of those modifications has to date been relatively limited. Based on time-stamps in ICL 2020c, the code has experienced, on average, average annual change traffic (number-of-statements-of-software-changed/total-number-software-statements in the system) of ~5%. This fraction is typical of software that has undergone relatively minor modifications, not of software that has been wholly re-engineered (Boehm 1981, 543; Boehm, Abts, Brown, Chulani, Clark, Horowitz et al. 2000, 28). This assessment was based on manually analyzing the source code in ICL 2020c and our experience with software engineering standards and practices.

### 5.1.5 Test Phase

This phase determines whether the product generated in the Implementation phase (Section 2.5) satisfies all requirements allocated to the software. Testing is typically performed at various software-build levels. There is no publicly accessible Test Plan, Test Report, or official Regression Test for ICL 2020c. (ICL 2020c does contain some test files, but what quality-control role those files are intended to support is not identified in ICL2020c). Eglen 2020 reports the results of porting, without modification, the source code in ICL 2020c to two small supercomputing platforms. Using test files provided by the ICL covid-19 simulator team (it is not clear these are the test files included in ICL 2020c), the ported code produced results that were the "same" as the results of some tables in ICL 2020b ("Report 9").  It should be noted, as Eglen 2020 does, that these demonstrations show nothing about the correctness of ICL 2020c.

### 5.1.6  Maintenance Phase

This phase iterates the phases described in Sections 2.2 – 2.6 after the product is deployed, as needed.  Maintenance policies and procedures are documented in a Maintenance Manual. There is no publicly accessible Maintenance Manual for ICL 2020c.

## 6.0  Discussion and conclusions

In this paper, we have argued that epidemiological simulators are de facto integral to epidemiological policy making. Justifiable policy-making in epidemiological crises such as the current COVID-19 pandemic involves trades among diverse values. Some of these values, such as tradeoffs between the rights of children and the rights of the elderly, lie outside the scope of epidemiology proper. Some of the values, including the need to assess the objective effects of various interventions, clearly lie within the scope of traditional epistemology. Furthermore, normative considerations play a role in determining what counts as an epidemic and what counts as an acceptable form of public health intervention.

We have explained why the norms from high-risk engineering contexts should be adopted in epidemiological contexts that have direct and significant public policy implications. In all projects of this kind, we urge teams to adopt methods that support transparency, explainability, and reproducibility within the framework of consensus safety-critical software engineering standards.  We urge journals and funding agencies to require that published results include access to a baseline instance of relevant software along with all the relevant documentation for implementation in order to ensure reproducibility and transparency.

Our assessment shows that ICL 2020c does not satisfy the standards for safety-critical software identified above.  It is clear, however, that the developers/maintainers of ICL 2020c have recently been modifying that simulator in a way that accommodates at least some of the consensus software engineering standards that we recommend.  This is commendable and we encourage the ICL team to continue work in this direction.